\documentclass{svmult}

\usepackage{makeidx}         
\usepackage{graphicx}        
\usepackage{amssymb}
\usepackage{amsmath}

\usepackage{multicol}        
\usepackage[bottom]{footmisc}

\spnewtheorem*{algorithm}{Proof of proposition}{\it}{\rm}
\newcommand{\Imm}{\mathrm{Im}}
\newcommand{\Inn}{Inn}
\newcommand{\Ker}{\mathrm{Ker}}
\newcommand{\Sym}{\mathrm{Sym}}
\newcommand{\F}{{\cal F}}
\newcommand{\G}{{\cal G}}
\newcommand{\A}{{\cal A}}
\newcommand{\Q}{\mathcal{P}}
\newcommand{\QQ}{\mathcal{Q}}
\newcommand{\RR}{\mathcal{R}}
\newcommand{\RH}{\mathcal{R}_{GL(4|3)}}
\renewcommand{\S}{\mathcal{S}}
\newcommand{\SH}{\mathcal{S}_{GL(4|3)}}
\newcommand{\W}{{\cal W}}
\renewcommand{\O}{{\cal O}}
\newcommand{\CC}{{\cal C}_{\mathbb{C}}}
\newcommand{\CR}{{\cal C}_{\mathbb{R}}}
\newcommand{\CH}{{\cal C}_{\mathbb{H}}}

\newcommand{\ity}{_{\infty}}
\newcommand{\Ber}{Ber((T^{\mathbb{C}}/\overline{\F})^*)}
\newcommand{\barF}{\overline{{\cal F}}}
\newcommand{\dbar}{\bar \partial}
\newcommand{\pr}[1]{\frac{\partial}{\partial #1}}
\newcommand{\mat}[4]{ \left(\begin{array}{cc} #1 & #2  \\ #3 & #4  \end{array} \right)}
\newcommand{\dirac}{D\!\!\!\!/}

\makeindex

\begin{document}

\title{Yang-Mills theory and a superquadric.}
\author{Michael Movshev\inst{1}}
\institute{Institute for Advanced Study \\ Einstein Dr 1, Princeton, NJ,USA
\\\texttt{mmovshev@ias.edu}}
\date{\today}

\maketitle
\begin{center}
\vskip .2in
{\it Dedicated to Yu.I. Manin on his 70-th birthday}
\vskip  .2in
\end{center}
\begin{abstract}
We construct a real-analytic CR supermanifold $\RR $, holomorphically embedded into a  superquadric $\QQ \subset \mathbf{P}^{3|3}\times \mathbf{P}^{*3|3}$. A CR distribution $\F$ on $\RR $ enables us to define a tangential CR complex $(\Omega_{\F}^{\bullet},\dbar)$.

We define  a $\dbar$-closed trace functional $\int:\Omega^{\bullet}_{\F}\rightarrow \mathbb{C}$ and  conjecture that a Chern-Simons theory associated with a triple $(\Omega^{\bullet}_{\F}\otimes Mat_n,\dbar,\int)$ is equivalent to N=3, D=4 Yang-Mills theory with a gauge group U$(n)$. We give some evidences to this conjecture. 

\end{abstract}
\section{Introduction}

Twistor  methods in  gauge theory   have a long history (summarized in ref. [P90]). A common  feature of these methods is that spacetime is replaced by a  twistor (or ambitwistor) analytic  manifold  $\mathbf{T}$ . Equations of motion "emerge" (in terminology of Penrose) from complex geometry of $\mathbf{T}$. 

The twistor approach turns to be a very useful technical innovation. For example difficult questions  of classical gauge theory, e.g. the ones that appear in  the theory of instantons,  admit a translation into a considerably more simple questions  of analytic  geometry of space  $\mathbf{T}$. On this  way classifications theorems in  the theory of instantons has been obtained (see ref. [AHDM]).

The quantum theory  did not have a simple reformulation  in the language of  geometry of the  space $\mathbf{T}$ so far. One of the  reasons is that the quantum theory formulated formally in terms of a path integral requires a Lagrangian. Classical theory, as it was mentioned earlier, provides only equations of motion whose definition  needs  no metric. In contrast a typical Lagrangian requires a metric in order to be defined. Thus a task of finding the Lagrangian in (ambi)twistor setup is not straightforward. In this paper we present a Lagrangian for N=3 D=4 Yang-Mills (YM) theory formulated in terms of  ambitwistors.

Recall that N=3 YM theory  coincides in components with N=4 YM theory. The easiest way to obtain N=4 theory is from N=1 D=10 YM theory by dimensional reduction. The Lagrangian of this ten dimensional theory is equal to 
\begin{equation}
(<F_{ij},F_{ij}>+<\dirac \chi,\chi>)dvol
\end{equation}
In the last formula $F_{ij}$ is a curvature of connection $\nabla$ in a principal U$(n)$-bundle over $\mathbb{R}^{10}$. An odd field  $\chi$ is a section of $S\otimes Ad$, where $S$ is a complex sixteen dimensional spinor bundle, $Ad$ is the adjoint bundle, $\dirac$ is the Dirac operator, $<.,.>$ is a Killing pairing on $\mathfrak{u}(n)$. The measure $dvol$ is associated with a flat Riemannian metric on $\mathbb{R}^{10}$, $F_{ij}$ are coefficients of the curvature in global orthonormal coordinates. The N=4   theory is obtained from this by considering fields invariant with respect to translations in six independent directions. The theory is conformally invariant and can be defined on any conformally flat manifold,e.g.  $S^4$ with a round metric. 

E. Witten in 1978 in ref.  [W78] discovered that it is possible to encode solutions N=3 supersymmetric YM-equation  by holomorphic  structures on a vector bundle defined over an open subset $U$ in a superquadric $\QQ$. We shall call the latter a complex ambitwistor superspace. In this description the action of N=3 superconformal symmetry on the space of solutions is manifest. Symbol $n|m$ denotes dimension of a supermanifold. More precisely the quadric  $\QQ \subset \mathbf{P}^{3|3}\times \mathbf{P}^{*3|3}$ is  defined by equation 

\begin{equation}\label{E:kdsj}
\sum_{i=0}^3 x_ix^i+\sum_{i=1}^3\psi_i\psi^i=0,
\end{equation}

 in bihomogeneous coordinates
\begin{equation}
x_0,x_1,x_2,x_3,\psi_1,\psi_2,\psi_3;\ x^0, x^1, x^2, x^3,\psi^1,\psi^2,\psi^3
\end{equation}

  in $\mathbf{P}^{3|3}\times \mathbf{P}^{*3|3}$($x_i,x^j$-even, $\psi_i,\psi^j$-odd, a symbol $*$ in the superscript stands for the dual space). The quadric a is complex supermanifold. It makes sense therefore to talk about differential $(p,q)$-forms $\Omega^{p,q}(\QQ)$.

Let $\G$ be a holomorphic vector bundle on $U$. Denote by
\begin{equation}\label{E:hjdur}
\Omega^{0\bullet}End\G,
\end{equation}
 a differential graded algebra  of smooth sections of $End\G$ with coefficients in $0,p$-forms.

Let $\dbar$ and  $\dbar'$ be two operators corresponding to two holomorphic structures in $\G$. It is easy to see that $(\dbar'-\dbar)b=ab$, where $a\in \Omega^{0,1}End\G$. The integrability condition $\dbar^{'^2}=0$ in terms of $\dbar$ and $a$ becomes a Maurer-Cartan (MC) equation:
\begin{equation}\label{E:fdsaf}
\dbar a+\frac{1}{2}\{a,a\}=0
\end{equation}

 The first guess would be that the space of fields of  ambitwistor version of N=3 YM would  be  $\Omega^{0,1}(U)End\G$, where $\G$ is a vector bundle on $U$ of some topological type. Witten  suggested in ref. [W03] that the  Lagrangian in question should be   similar to a  Lagrangian of holomorphic Chern-Simons theory.

\begin{equation}\label{E:gdfsguqd}
CS(a)=\int tr(\tilde(\frac{1}{2}a\dbar a +\frac{1}{6}a^3))Vol
\end{equation}
where $Vol$ is some integral form. The action (\ref{E:gdfsguqd}) reproduces equations of motion (\ref{E:fdsaf}). The hope is that perturbative analysis of this quantum theory will give some insights on the structure of N=3 YM.

The main result of the present note is that we 
give a precise meaning to this conjecture.

Introduce a real supermanifold $\RR \subset \QQ$ of real superdimension $8|12$. It is defined by equation 
\begin{equation}\label{E:dfsad}
x_1\bar{x}^2-x_2\bar{x}^1+x_3\bar{x}^4-x_4\bar{x}^3+\sum_{i=1}^3\psi_i\bar{\psi}^i=0
\end{equation}
In Section \ref{S:llfddfgg} we discuss the meaning of reality in superalgebra and geometry.
\begin{definition}
Let $M$ be a $C^{\infty}$ supermanifold, equipped with a subbundle $H$ of the tangent bundle $T$. We say that $M$ is equipped with CR-structure if $H$ carries a complex structure defined by fiberwise transformation $J$.

The operator $J$ defines a decomposition of the complexification $H^{\mathbb{C}}$ into a direct sum of eigensubbundles $\F+\barF$.

We say that the CR-structure $J$ is integrable if the sections of $\F$ form a Lie subalgebra of $T^{\mathbb{C}}$ under the bracket of vector fields.
\end{definition}

A tautological  embedding of $\RR $ into the complex manifold $\QQ$ induces a CR structure specified by distribution $\F$. Properties of this CR structure are discussed in Section \ref{S:qjscmmz}. A global holomorphic supervolume form $vol$ on $\QQ$  is constructed in Proposition \ref{P:jsdhd}. When restricted on $\RR $ it defines a section of $\ ^{int}\Omega^{-3}_{\F}$-a CR integral form. Functorial properties of this form are discussed in Section \ref{S:gufjasklly}. For any CR-holomorphic vector bundle $\G$ we define a differential graded algebra $\Omega^{\bullet}_{\F}End(\G)$. It is the tangential CR complex. We equip it with a trace 
\begin{equation}
\int :a \rightarrow \int_{\RR }tr(a)\ vol
\end{equation}
We define a CS-action of the form (\ref{E:gdfsguqd}), where we replace an element of $\Omega^{0\bullet}(U)End(\G)$ by an element of $\Omega^{\bullet}_{\F}(\RR )End(\G)$. The integral is taken with respect to the  measure $vol$. We make some assumptions about topology of $\G$ as it is done in classical twistor theory. The space   $\S$ is a superextension of $S^4$ (see Section \ref{S:hfdgjpo} for details). There is a projection 
\begin{equation}\label{E:sdsfsew}
p:\RR \rightarrow \S
\end{equation}
 We require that $\G$ is topologically trivial along the fibers of $p$. It is an easy exercise in algebraic topology to see that topologically all such bundles are pullbacks from $S^4$. On $S^4$ unitary vector bundles are classified by their second Chern classes. 
\begin{conjecture}\label{C:gdfsgjj}
Suppose $\G$ is a CR-holomorphic vector bundle on $\RR $ of rank $n$.
Under the above assumptions a  CS theory defined by the algebra $\Omega^{\bullet}_{\F}End(\G)$ is equivalent to N=3 YM theory on $S^4$ in a principal U$(n)$ bundle   with the  second Chern class equal to $c_2(End(\G))$.
\end{conjecture}

For perturbative computations in YM theory it is convenient to work in BV formalism. See ref. [Sch00] for mathematical introduction and ref. [MSch06] for applications to YM. 
\begin{conjecture}\label{C:ttc2}
In assumptions of Conjecture \ref{C:gdfsgjj} we believe that N=3 YM theory in BV formulation is equivalent to a CS theory defined by the algebra $\Omega^{\bullet}_{\F}End(\G)$,  where the field  $a\in \Omega^{\bullet}_{\F}End(\G)$ has a mixed degree.
\end{conjecture}

The following abstract definition will be useful
\begin{definition}\label{D:rtje}
Suppose we are given a differential graded algebra $(A,d)$ with a $d$-closed trace functional $\int$. We can consider $A$ as a space of fields in some field theory with Lagrangian defined by the formula
\begin{equation}
CS(a)=\int(\frac{1}{2}ad(a)+\frac{1}{6}a^3)
\end{equation}
We call it a Chern-Simons $CS$ theory associated with a triple $(A,d,\int)$

We say that two theories $(A,d,\int)$ and $(A',d',\int')$ are classically equivalent if there is a quasiisomorphism of algebras with trace $f:(A,d,\int)\rightarrow (A',d',\int')$

See Appendix of [MSch05] for extension of this definition on A$\ity$ algebras with a trace.
\end{definition}

Thus the matrix-valued Dolbeault complex $(\Omega_{\F}^{\bullet}(\RR)\otimes Mat_n,\dbar)$ with a trace defined by the formula $\int(a)=tr\int_ravol$ would give an example of such algebra.

We shall indicate existence of classical equivalence of N=3 YM defined over $\Sigma=\mathbb{R}^4\subset S^4$ and a CS theory defined over $p^{-1}(U)$,

where $U$ is an open submanifold of $\S$ with $U_{red}=\Sigma$.

Here is the idea of the proof.

We  produce a supermanifold $Z$ and  an integral form $Vol$ on it. We  show that a CS theory constructed using   differential graded algebra with a trace $A(Z)$, associated with   manifold $Z$ is classically equivalent to N=3 YM theory.  We  interpret algebra $A(Z)$ as a tangential CR-complex on $Z$.

We shall construct a manifold $Z$ and algebra $A$ in two steps .

Here is a description of the steps in more details:

{\bf Step 1.} We define a compact analytic supermanifold $\widetilde {\Pi F}$ and construct  an integral form $\mu$ on it in a spirit of [MSch05]. 
 Let $A_{pt}$ be the Dolbeault complex of $\widetilde {\Pi F}$. Integration of an element $a \in A_{pt}$ against $\mu$ over $\widetilde {\Pi F}$ defines a $\dbar$-closed trace functional on $A_{pt}$. We show that the Chern-Simons theory associated with dga $(A_{pt}\otimes Mat_n,\dbar,\int\otimes tr_{Mat_n})$ is classically equivalent to N=3 Yang-Mills theory with gauge group U$(n)$ reduced to a point.

{\bf Step 2.} From the algebra $A_{pt}$ we reconstruct a differential algebra $A$. The algebra $A\otimes Mat_n$ conjecturally encodes full N=3 Yang-Mills theory with gauge group U$(n)$ in a sense of Definition  \ref{D:rtje}. If we put aside the differential $d$,  $A$  is equal to $A_{pt}\otimes C^{\infty}(\Sigma)$. The integral form we are looking for is equal to $Vol=\mu dx_1dx_2dx_3dx_4$, where $\Sigma$ is equipped with global coordinates $x_1,x_2,x_3,x_4$.

The manifold $Z$ is a CR submanifold. We identify it with an open subset of $\RR $.

Finally we would like to formulate an unresolved question. Restriction of a holomorphic vector bundle $\G$ over $U$ on $U\cap \RR$ defines a CR-vector bundle over the intersection. Is it true that any CR-holomorphic vector bundles can be obtained this way? The answer would be affirmative if we impose some analyticity conditions on the CR structure on $\G$. Presumably super Levi form will play a role in a solution of this problem.

It is tempting to speculate that there is a string theory on $\QQ$ and $\RR $ defines  a D-brane in it.
 
We need to say few words about the structure of this note. 
In Section  \ref{S:dfdrf} we make definitions and provide some constructions used in formulation of conjectures (\ref{C:gdfsgjj}) and (\ref{C:ttc2}).

 In Section \ref{S:main} we give a geometric twistor-like description of N=3  YM theory reduced to a point( Step 1). In Section \ref{S:ffkewx} we do the Step 2.

Appendix contains some useful definitions concerning reality in superalgebra and CR-structures.

\section{Infinitesimal constructions}\label{S:dfdrf}
In  this section we shall show that the space $\RR $ is  homogeneous with respect to the action of a real form of N=3 superconformal algebra $\mathfrak{gl}(4|3)$ . Here we also collected  facts that are needed for coordinate-free description of  space $\RR$ in terms of Lie algebras of symmetry group and isotropy subgroup.

\subsection{Real structure on the Lie algebra $\mathfrak{gl}(4|3)$}\label{S:qjscmmz}

In this section we describe a graded real structure on $\mathfrak{gl}(4|3)$.  It will be used later in construction of CR-structure on real super-ambitwistor space.

The reader might wish to consult Section \ref{S:llfddfgg} for the definition of a graded real  structure. There the reader will find explanation of some of our notations. By definition  $\mathfrak{gl}(4|3)$ is a super Lie algebra of endomorphisms of  $\mathbb{C}^{4|3}=\mathbb{C}^{4}+\Pi\mathbb{C}^{3}$. Symbol $\Pi$ stands for the parity change.  This Lie algebra  consists of matrices of a block form $\mat{A}{B}{C}{D}$  with $A \in Mat(4 \times 4,\mathbb{C}),D \in Mat(3\times 3,\mathbb{C}), C \in Mat(3\times 4,\mathbb{C}),B\in Mat(4 \times 3,\mathbb{C})$. Elements $\mat{A}{0}{0}{D}$ belong the even part $\mathfrak{gl}_{0}(4|3)$, elements $\mat{0}{B}{C}{0}$  to the odd  $\mathfrak{gl}_{1}(4|3)$.

In the following a symbol $\mathfrak{g}(\mathbb{K})$ will stand for a Lie algebra defined over a field $\mathbb{K}$. If the field is not present it means that the algebra is defined over $\mathbb{C}$. The same applies to Lie groups.

Let $\mathfrak{g}$ be a complex super Lie algebra. By definition a map $\rho$ that  defines a graded real structure on on super Lie algebra $\mathfrak{g}$   if $\rho$ is a homomorphism: $\rho[a,b]=[\rho(a),\rho(b)]$. In ref. [Man] Yu. I. Manin suggested several definitions of a real structure on a (Lie) superalgebra. In notations of [Man] these definitions are parametrized by  a triple $(\epsilon_1,\epsilon_2,\epsilon_3), \epsilon_i=\pm$. Our real structure correspond to the choice  $\epsilon_1=-,\epsilon_2=\epsilon_3=+$.

The reader will find a complete classification of graded real structures on simple Lie algebras in  the work [Serg].

Define a  matrix $J$ as :
\begin{equation}
J=\mat{0}{id}{-id}{0}
\end{equation}
where $id$ is a $2\times 2$ identity matrix. A map $\rho$ is defined as 

\begin{equation}\label{E:gfjeiuq}
\rho\mat{A}{B}{C}{D}=\mat{J}{0}{0}{id}\mat{\overline{A}}{\overline{B}}{\overline{C}}{\overline{D}}\mat{-J}{0}{0}{id}=\mat{-J\overline{A}J}{J\overline{B}}{-\overline{C}J}{\overline{D}}
\end{equation}
The identity $\rho^2=sid$\footnote{The operator $sid$ is defined in the Appendix in Definition \ref{D:gdfjgdf}.} is a corollary of equation $J^2=-id$.

It is useful to analyze the Lie subalgebra $\mathfrak{gl}_{0}(4|3)^{\rho}$ of real points  in  $\mathfrak{gl}_0(4|3)=\mathfrak{gl}(4,\mathbb{C})\times \mathfrak{gl}(3,\mathbb{C})$. Due to (\ref{E:gfjeiuq}) we have  $\mathfrak{gl}(3)^{\rho}=\mathfrak{gl}(3,\mathbb{R})$. To identify $\mathfrak{gl}^{\rho}(4)$ we interpret $\mathbb{C}^{4}=\mathbb{C}^{2}+ \mathbb{C}^{2}$(whose  algebra of endomorphisms is $\mathfrak{gl}(4)$ ) as a  two-dimensional quaternionic space $\mathbb{H}+\mathbb{H}$. Let $1,i,j,k$ be the standard $\mathbb{R}$-basis in quaternions, $<e_1,e_2>$ be an  $\mathbb{H}$-basis in $\mathbb{H}+\mathbb{H}$. The space $\mathbb{H}+\mathbb{H}=\mathbb{C}^{2}+ \mathbb{C}^{2}$  has a complex structure defined by the  right multiplication on $i$. The right multiplication on $j$ defines an $i$-antilinear map. In a $\mathbb{C}$-basis $e_1,e_2,e_1j,e_2j$ a matrix of right multiplication on $j$ is equal to $J$. From this it is straightforward to deduce that $\mathfrak{gl}^{\rho}(4)=\mathfrak{gl}(2,\mathbb{H})$.

\begin{definition}
Let $M$ be a $C^{\infty}$ supermanifold with a tangent bundle $T$. Let $H\subset T$ be a subbundle equipped with a complex structure $J$. This data defines a (nonintegrable) CR-structure on $M$. There is a  decomposition \footnote{In the following a letter $\mathbb{C}$ in superscript denotes complexification.} $H^{\mathbb{C}}=\F+\overline{\F}$. A CR-structure $(H,J)$ is integrable if a space of sections of $\F$ is closed under commutator. In this case we also  say that $\F$ is integrable.
\end{definition}
\begin{definition}
Let $M_{red}$ denote the underlying manifold of supermanifold $M$.
\end{definition}

If $M$ is a real submanifold  of a complex supermanifold $N$ then at any $x\in M$ the tangent space $T_x$ to $M$ contains a maximal complex subspace $H_x$. If $rank H_x$ is constant along $M$ then a family of spaces $H$ defines an integrable CR-structure. In our case the manifold $\RR \subset \QQ$ is defined by equation (\ref{E:dfsad}).

Denote by $GL(4|3)$ an affine supergroup with Lie algebra $Lie(GL(4|3))$ equal to $\mathfrak{gl}(4|3)$\footnote{For global description of $(GL(4|3),\rho)$ see ref. [Pel].}. We will show later that $\RR$ is a homogeneous space of a real form of $GL(4|3)$ described above. The induced CR-structure is real-analytic and homogeneous with respect to the group action . 

A CR-structure on a supermanifold enables us to define an analog of Dolbeault complex . Suppose a supermanifold $M$ carries a CR structure  $\F\subset T^{\mathbb{C}}$. A space of complex 1-forms $\Omega^{1}_M$ contains a subspace $I$ of forms pointwise orthogonal to $\F$. It is easy to see that  $\F$ is integrable iff the ideal $(I)$ is closed under $d$. Define a tangential CR-complex $(\Omega^{\bullet}_{\F},\dbar)$ to be $(\Omega^{\bullet}/(I),d)$ .

A vector bundle $\G$ is CR-holomorphic if the gluing cocycle $g_{ij}$ satisfies $\dbar g_{ij}=0$. In such case we can define a $\G$-twisted CR-complex  $\Omega_{\F}^{\bullet}\G$.

\begin{remark}
Denote by $\sigma$ an operation of complex conjugation. Define an antilinear map 
\begin{equation}\label{E:gjhsirpqqa}
s=\sigma\circ\mat{J}{0}{0}{id}:\mathbb{C}^{4|3}\rightarrow \mathbb{C}^{4|3} .
\end{equation}
 A map $a\rightarrow sas^{-1}, a\in \mathfrak{gl}(4|3)$ coincides with the real structure $\rho$. Let us think about LHS of equation (\ref{E:kdsj}) as  a quadratic function associated with an even bilinear form $(a,b)$. It is easy to see that LHS of equation (\ref{E:dfsad}) is equal to  $(a,s(a))=0$.   Naively thinking the centralizer of  operator $s$ would  precisely be the real form of $(\mathfrak{gl}(4|3), \rho)$ and it would preserve equations (\ref{E:kdsj}) and (\ref{E:dfsad}). The problem is that we cannot work pointwise in supergeometry. Instead we consider equations (\ref{E:kdsj}, \ref{E:dfsad}) as a system of real algebraic equations. We interpret them as a system of sections of some line bundles on $\CH$ manifold $M=\mathbf{P}^{3|3}\times \mathbf{P}^{*3|3}\times \overline{\mathbf{P}}^{3|3}\times \overline{\mathbf{P}}^{*3|3}$ (see Section \ref{S:llfddfgg} for discussion of reality in supergeometry). The space $M$ carries a canonical graded real structure $\rho$, that leaves the space of equations invariant. The  $\rho$-twisted diagonal action of $\mathfrak{gl}(4|3)$ also leaves the equations invariant. 

The graded real structure  induces a graded real structure $\rho$ on $\mathfrak{gl}(4|3)^{\rho}$ and makes a supermanifold $\RR $ an algebraic graded real  supermanifold.

\end{remark}

\subsection{Symmetries of the ambitwistor space}\label{S:hfdgjpo}
We define a space $\RH $ as  a homogeneous space of a real supergroup $(GL(4|3),\rho)$. In this section we establish an isomorphism $\RH\cong \RR$.

\begin{figure} [ht]
\centering
\includegraphics[width=.3\textwidth,height=.3\textwidth]{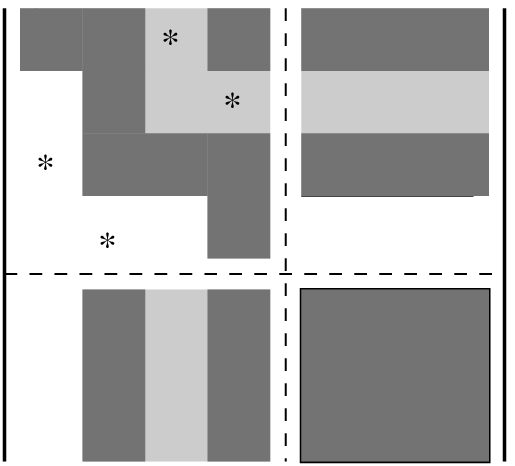}
\caption{}
\label{F:gdfgdjh}
\end{figure}

In fig. (\ref{F:gdfgdjh}) the reader can see a graphical presentation of some matrix $\mat{A}{B}{C}{D}\in \mathfrak{gl}(4|3)$.

The isotropy subalgebra $\mathfrak{a}\subset \mathfrak{gl}(4|3)$ of a base point in the space $\RH $ is defined as a linear space of matrices whose nonzero entries are in the darkest shaded area of a matrix  in fig. (\ref{F:gdfgdjh}).
\begin{lemma}
Subspace $\mathfrak{a}\subset \mathfrak{gl}(4|3)$ is a $\rho$-invariant subalgebra.
\end{lemma}
\begin{proof}
Direct inspection.
\end{proof}
Let $A$ be an algebraic subgroup of $GL(4|3)$ with a Lie algebra $\mathfrak{a}$.

The space $\RH $ carries a homogeneous CR structure (see Section  \ref{S:gfgjheq} for related discussion). Define a subspace $\mathfrak{p}\subset \mathfrak{gl}(4|3)$ as a set of matrices with nonzero entries in gray and  dark gray  areas in fig. (\ref{F:gdfgdjh}).
\begin{lemma}
Subspace $\mathfrak{p}\subset \mathfrak{gl}(4|3)$ is a subalgebra. It satisfies $\mathfrak{p}\cap \rho(\mathfrak{p})=\mathfrak{a}$
\end{lemma} 
\begin{proof}
Direct inspection.
\end{proof}
Let $P$ denote an algebraic subgroup with Lie algebra  $\mathfrak{p}$ .
A complex supermanifold $X=GL(4|3)/P$ has  an explicit description.

Equation (\ref{E:kdsj}) is preserved by the  action of $GL(4|3)$. 
\begin{proposition}
There is a $GL(4|3)$-equivariant isomorphism $X=\QQ$.
\end{proposition}
\begin{proof}

 We can identify the quadric $\QQ$ with the space of partial flags ${\mathbb C}^{4|3}$ in as it is done in purely even case (see [GH] for example). A spaces $\QQ$ is  a connected component of the flag space containing the  flag 
\begin{equation}\label{E:yhggb}
F_1\subset F_2\subset {\mathbb C}^{4|3}
\end{equation}
with $F_1\cong {\mathbb C}^{1|0}$ and $F_2\cong {\mathbb C}^{3|3}$. This flag  can be interpreted as  a pair of points $F_1\in \mathbf{ P}^{3|3}$, $F_2\in \mathbf{P}^{*3|3}$.  The condition (\ref{E:yhggb}) is equivalent to (\ref{E:kdsj}).

Let us choose a standard basis $e_1,\dots,e_7$ of ${\mathbb C}^{4|3}$ such that  the parities of elements are  $\varepsilon(e_1)=\varepsilon(e_2)=\varepsilon(e_3)=\varepsilon(e_7)=1$, $\varepsilon(e_4)=\varepsilon(e_5)=\varepsilon(e_6)=-1$. In this notations the standard flag $F$ has the following description:
\begin{equation}\label{E:ksk}
\begin{split}
&F_1=\mbox{span}<e_7>\\
&F_2=\mbox{span}<e_2,\dots e_7>
\end{split}
\end{equation}
The flag defines a point in the space $\QQ$.
It is easy to compute  a shape of the   matrix of an element from the stabilizer $P_{F}$ of $F$. The following picture is useful 
\begin{equation}\label{E:fgojkxcz}
\mbox{ 
\setlength{\unitlength}{3947sp}%
\begin{picture}(2744,2456)(151,-1573)
\thicklines
\put(226,389){\line( 1, 0){600}}
\put(826,389){\line( 0,-1){1500}}
\put(826,-1111){\line( 1, 0){1575}}
\put(2401,-1111){\line( 0,-1){375}}
\thinlines
\multiput(826,-1111)(0.00000,-120.00000){3}{\line( 0,-1){ 60.000}}
\multiput(826,-1411)(120.00000,0.00000){13}{\line( 1, 0){ 60.000}}
\multiput(2326,-1411)(0.00000,120.00000){3}{\line( 0, 1){ 60.000}}
\multiput(2326,-1111)(-120.00000,0.00000){13}{\line(-1, 0){ 60.000}}
\multiput(376,314)(0.00000,-123.91304){12}{\line( 0,-1){ 61.957}}
\multiput(376,-1111)(107.14286,0.00000){4}{\line( 1, 0){ 53.571}}
\multiput(751,-1111)(0.00000,123.91304){12}{\line( 0, 1){ 61.957}}
\multiput(751,314)(-107.14286,0.00000){4}{\line(-1, 0){ 53.571}}
\thicklines
\multiput(826,839)(0.00000,-300.00000){2}{\line( 0,-1){150.000}}
\multiput(826,389)(286.36364,0.00000){6}{\line( 1, 0){143.182}}
\multiput(2401,389)(0.00000,-272.72727){6}{\line( 0,-1){136.364}}
\multiput(2401,-1111)(300.00000,0.00000){2}{\line( 1, 0){150.000}}
\thinlines
\put(2401,839){\line( 1, 0){450}}
\put(2851,839){\line( 0,-1){2400}}
\put(2851,-1561){\line(-1, 0){375}}
\thicklines
\put(2776,-1486){\line(-1, 0){375}}
\put(2401,-1486){\line( 0, 1){  0}}
\put(2401,-1486){\line( 0, 1){  0}}
\thinlines
\put(676,839){\line(-1, 0){450}}
\put(226,839){\line( 0,-1){2400}}
\put(226,-1561){\line( 1, 0){375}}
\put(1276,-436){\makebox(0,0)[lb]{$\mathfrak {gl}(2|3)$}}%
\put(1276,-1336){\makebox(0,0)[lb]{$\mathbb{C}^{*2|3}$}}%
\put(2476,-1336){\makebox(0,0)[lb]{$\mathbb {C}$}}%
\put(451,539){\makebox(0,0)[lb]{$\mathbb{C}$}}%
\put(2401,539){\makebox(0,0)[lb]{$\mathfrak{rad}$}}%
\put(451,-1336){\makebox(0,0)[lb]{$\mathbb{C}^{1|0}$}}%
\put(451,-436){\makebox(0,0)[lb]{$\mathbb{C}^{2|3}$}}%
\end{picture}
}
\end{equation}
The  Lie algebra $\mathfrak{p}_F$ of stabilizer $P_F$ is formed by matrices with zero entries below thick solid line in  picture (\ref{E:fgojkxcz}).
Conjugating with a suitable permutation of coordinates $t$ we see that $\mathfrak{p}^t_F=\mathfrak{p}$.
\end{proof}

\begin{remark}
A transformation $s$ defined in equation (\ref{E:gjhsirpqqa})  acts on the space of flags.
By definition an $s$-invariant   flag  belongs to the subvariety $\RR $ .
A direct inspection shows that the nonzero entries of the matrix of an element of stabilizer are located in the darkest shaded area  on  fig. (\ref{F:gdfgdjh}). The manifold $\RR _{red}$ fibers over $\mathbf{P}^3$ with connected fibers. Thus $\RR $ is connected. From this and a simple dimension count we conclude that subvariety $\RR $  coincides with $\RH $. Identification of CR structure also follows from this.
\end{remark}
The Lie algebra $\mathfrak{gl}(4|3)$ contains a subalgebra $\mathfrak{l}$. The elements of this subalgebra have nonzero entries in the darkest area on fig (\ref{F:gdfgdjh}) and also spots marked by $*$. This algebra is invariant with respect to the real structure $\rho$. Denote by $L$ an algebraic subgroup of $GL(4|3)$ with Lie algebra $\mathfrak{l}$. 

The quotient $(\SH,\rho)=((GL(4|3)/L),\rho)$ is a supermanifold with $(\SH)^{\rho}_{red}=S^4$. Indeed  the real points $L_{red}^{\rho}$ of the group $L_{red}$ are conjugated to quaternionic matrices of the form $\mat{a}{b}{0}{d}\in GL(2,\mathbb{H})$. Thus the quotient space $GL(2,\mathbb{H})/L^{\rho}_{red}=\mathbb{H}\mathbf{P}^1$ is isomorphic to $S^4$. Denote by $p$ projection 
\begin{equation}\label{E:sdjhwuwq}
\RH \rightarrow \SH
\end{equation}
An easy local exercise with Lie algebras reveals that the fibers of projection $p$ are CR-holomorphic and are isomorphic to $\mathbf{P}^1\times\mathbf{P}^1$.

The following direct geometric description  of ambitwistor space will be useful. Let $M$ be a $C^{\infty}$ 4-dimensional Riemannian manifold. A metric $g$ defines a relative quadric (the ambitwistor space) in the projectivisation of a complexified tangent bundle $A(M)\subset \mathbf{P}(T^{\mathbb{C}})$. By construction there is a projection $p:A(M)\rightarrow M$. The space $A(M)$ carries a  CR structure (it could be nonintegrable). Indeed a fiber of the distribution $\F$ at a point $x\in A(M)$ is a direct sum of the holomorphic tangent space to the  fiber through $x$ and a complex line in  $T^{\mathbb{C}}(M)$ spanned by $x$. From the point of view of topology the space $A(M)$ coincides with a relative Grassmannian of oriented 2-planes in $T_M$. A constructed complex distribution depends only on conformal class of the metric. From this we conclude that $A(S^4)$ ($S^4$ has a round metric induced by the standard embedding into $\mathbb{R}^5$) is a homogeneous space of $Conf(S^4)=PGL(2,\mathbb{H})$, $A(S^4)=PGL(2,\mathbb{H})/A_{red}^{\rho}$ and  the CR structure is integrable.

An appropriate  super generalization of this construction is as follows . We have an isomorphism
\begin{equation}\label{E:aadfjs}
\W_l\otimes \W_r\overset{\Gamma}{\cong}T^{\mathbb{C}}_M
\end{equation} 
In the last formula $\W_l,\W_r$ are complex two-dimensional spinor bundles on $M$ (we assume that $M$ has a spinor structure). The isomorphism $\Gamma$ is defined by Clifford multiplication.  Let $T$ be   a 3-dimensional linear space. This vector space will enable us to implement N$=dim(T)=3$ supersymmetry. To simplify notations we keep  $\W_l\otimes T+\W_r\otimes T^*$ for the pullback  $p^*(\W_l\otimes T+\W_r\otimes T^*)$. Define a split, holomorphic in odd directions\footnote{The reader might wish to consult Section \ref{S:llfddfgg}  on this.} supermanifold $\widetilde{\A}(M)$ associated with a vector bundle  $\Pi(\W_l\otimes T+\W_r\otimes T^*)$ over $A(M)$. To complete the construction we  define a superextension of the CR structure. Introduce odd local coordinates $\theta^i_{\alpha},\tilde{\theta}^{j\beta}$ ($1\leq i,j\leq 2,1\leq \alpha,\beta\leq 3$) on fibers of $\Pi(\W_l\otimes T+\W_r\otimes T^*)$. We decompose  local complex vector fields $\Gamma(\pr{\theta^i_{\alpha}}\otimes \pr{\tilde{\theta}^{j\beta}})$ in a local real basis $\pr{x^s}$ as  $\delta_{\alpha}^{\beta}\Gamma^{ s}_{ij}\pr{x^s}, 1\leq s\leq 4$. The odd part of the CR distribution $\F$ is locally spanned by vector fields
\begin{equation}\label{E:gfdjfur}
\begin{split}
&\pr{\theta_{i\alpha}}+\tilde{\theta}^{j}_{\alpha}\Gamma_{ij}^{s}\pr{x^{s}}\\
& \pr{\tilde{\theta}^{j}_{\alpha}}+\theta^{j\alpha}\Gamma^{s}_{ij}\pr{x^{s}}
\end{split}
\end{equation}
This construction of a superextension  of ordinary CR structure  depends only on the conformal class  of the metric. It is convenient to formally add the complex conjugate odd coordinates. This way we get $\A(M)=\Pi(\W_l\otimes T+\W_r\otimes T^*+\overline{\W_l\otimes T}+\overline{\W_r\otimes T^*})$, equipped with a graded real structure. As in the even case the symmetry analysis allows to identify CR-space $\A(S^4)$ with $\RR $.

A tangent space $\mathfrak{ m}$ to $\QQ$ at a point fixed by $\mathfrak{p}$ is formed by elements with nonzero entries below thick solid line on the picture (\ref{E:fgojkxcz}) . It  decomposes into a sum ${\mathbb C}^{2|3}+{\mathbb C}^{*2|3}+{\mathbb C}^{1|0}$ of irreducible $GL(2|3)$ representations.

The elements $a_{21},a_{31},\alpha_{41},\alpha_{51},\alpha_{61},a_{71},a_{72},a_{73},\alpha_{74},\alpha_{75},\alpha_{76}$ stand for matrix coordinate functions  on the linear space $\mathfrak{ m}$( coordinates $a$ are even, $\alpha$ are odd).

\begin{proposition}\label{P:jsdhd}
An element
\begin{equation}\label{e:jksdss}
vol=da_{21}\wedge da_{31}\wedge d\alpha_{41}\wedge d\alpha_{51}\wedge d\alpha_{61}\wedge da_{71}\wedge da_{72}\wedge da_{73}\wedge d\alpha_{74}\wedge d\alpha_{75}\wedge d\alpha_{76}
\end{equation}
belongs to  $Ber(\mathfrak{ m}^*)$. It is invariant with respect to the action of $P$.
\end{proposition}
\begin{proof}
Simple weight count.

\end{proof}

We spread a generator of $Ber(\mathfrak{ m}^*)$ by the action of $GL(4|3)$ over $\QQ$ and form a $GL(4|3)$-invariant section $vol$ of $Ber_{\mathbb{C}}(\QQ)$

\section{Reduced theory}\label{S:main}
As a preliminary step in construction of the superspace $Z$ we introduce a "holomorphic" manifold $\widetilde{\Pi F} $ and an integral form on it (see Section \ref{E:gfgdfgjr}  for the explanation of quotation marks ). The form defines a functional  $\int$ on Dolbeault complex of this manifold. We prove that the  CS theory constructed by the triple $(\Omega^{0\bullet}(\widetilde{\Pi F})\otimes Mat_n,\dbar,\int\otimes tr_{Mat_n})$  is classically equivalent to  N=3 YM theory with gauge group U$(n)$ reduced to a point.

\subsection{ A Manifold $\Pi F$}\label{S:ncjsjs}

A manifold $\widetilde{\Pi F}$ is a deformation of a more simple manifold $\Pi F$. In this section we give relevant definitions  concerning $\Pi F$.

Denote a product $\mathbf{P}^1\times \mathbf{P}^1$ by $X$. It has two projections $p_i:X\rightarrow \mathbf{P}^1 , i=l,r$. Let ${\cal O}(1)$ denote the dual to the Hopf line bundle over $\mathbf{P}^1$. The Picard group of $X$ is $\mathbb{Z}+\mathbb{Z}$. It is generated by the classes of line bundles $\pi ^*_l{\cal O}(1)={\cal L}_l$, \quad$\pi ^*_r {\cal O}(1)={\cal L}_r$ that can serve as  coordinates in $Pic(X)$. Let ${\cal O}(a,b)$ denote a line bundle ${\cal L}^{\otimes a}_l\otimes {\cal L}^{\otimes b}_r$.  

{\bf Convention} We denote by $H^{\bullet}(Y,\G)$ the cohomology of (super)manifold  $Y$ with coefficients in a vector bundle $\G$. It can be computed as cohomology of Dolbeault complex $\Omega^{0\bullet}(Y)\G$. It is tacitly assumed  that  in  the Section \ref{S:main} the omitted argument $Y$ in $\Omega^{0\bullet}(Y)\G$ implies $Y=X$ . If $\G$-argument is missing we assume that $\G={\cal O}$.

Denote by  $\Sym V,\Lambda V$ symmetric and exterior algebras of a vector space (bundle).

Denote by $\Theta$ a line bundle isomorphic to ${\cal O}(1,1)$. 
We construct a vector bundle $F$ over $X$ as  a direct sum:
\begin{equation}\label{E:sjs}
\begin{split}
&F=T\otimes {\cal L}_l+T^*\otimes {\cal L}_r+\Theta^* \\
&H=T\otimes{\cal L}_l+T^*\otimes {\cal L}_r.
\end{split}
\end{equation} 
As before $T$ is a three dimensional  vector space.

The reader may have noticed that the manifold $X$  has also appeared as  a fiber of projection (\ref{E:sdsfsew}).  We shall see this is not accidental.
\subsection{Properties of the manifold $\Pi F$}\label{E:gfgdfgjr}

In this section we devise an infinitesimal deformation of a complex structure on $\Pi F$. This deformation will be promoted to the actual deformation which we denote  by $\widetilde{\Pi F}$. The algebra $A_{pt}$ from the introduction is equal to $\Omega^{0\bullet}(\widetilde{\Pi F})$. We construct on $\widetilde{\Pi F}$ we construct an integral form that will enable us to define a functional $\int_D:\Omega^{0\bullet}(\widetilde{\Pi F})\rightarrow \mathbb{C}$.

The manifold $\Pi F$ is a complex split supermanifold. The Dolbeault complex $(\Omega ^{0\bullet}(\Pi F),\dbar)$ is defined on supermanifold $\Pi F$, considered as a graded real supermanifold.  The complex $(\Omega ^{0\bullet}(\Pi F),\dbar)$ contains as a differential subalgebra the Dolbeault complex $\Omega ^{0\bullet}\Lambda F^*$. 

\begin{proposition}
Differential algebras $\Omega ^{0\bullet}(\Pi F)$ and $\Omega ^{0\bullet}\Lambda F^*$ are quasiisomorphic.
\end{proposition}
\begin{proof}
The same as a proof of Proposition \ref{E:fdsafjh}.
\end{proof}
The canonical line bundle ${\cal K}_X$ is equal to ${\cal O}(-2,-2)$. There is a nontrivial cohomology class- the "fundamental" class: $\alpha \in H^2(X,{\cal O}(-2,-2))\overset {id}{\subset } H^2(X,{\cal O}(-2,-2)\otimes T\otimes T^*)\subset H^2(X,\Lambda^2 (H^*)\otimes \Theta ^*)\subset H^2(X,\Lambda F^*\otimes \Theta^*)$. 
We interpret $\Lambda (F^*)\otimes \Theta^* $ as a sheaf of local holomorphic differentiations of $\Pi F$ in direction of $\Theta^*$. 

A representative $\alpha =f d\bar {z}_ld\bar{z}_r\pr{\theta}$ ($z_l,z_r$ are local coordinates on $X$) of the class $[\alpha ]$ can be extended to a  differentiation of $\Omega ^{0\bullet}(\Pi F)$.
The main properties of $D=\dbar+\alpha$ are:
\\
$1)$ it is a differentiation of $\Omega ^{0\bullet}(\Pi F)$, 
\\
$2)$ equation  $ D^2=0$ holds.
\\
All of them are corollaries of $\dbar $-cocycle equation for  $\alpha$.
The operator $D$ defines a new "holomorphic" structure on $\Pi F$ \footnote{This definition is not standard, because usually a deformation cocycle  $\alpha$  is an element of $ \Omega^{0,1}{\cal T}$ (${\cal T}$ is a holomorphic tangent bundle), whereas in our case $\alpha \in \Omega^{0,2}{\cal T}$. We however continue to use a traditional wording and call it a deformation of a complex structure, though a more precise term would be deformation of the Dolbeault algebra $(\Omega^{0\bullet},\bar)$.   This algebra in our approach becomes a substitute for  the underlying manifold. 

}. 
This new complex manifold will be denoted by $\widetilde{ \Pi F}$.

The manifold $\Pi F$ is Calabi-Yau. By this we mean that $Ber_{\mathbb{C}}$ is trivial. Indeed the determinant line bundle of $F$ is equal to $det(T\otimes {\cal L}_l)\otimes det (T^*\otimes {\cal L}_r)\otimes det ({\cal O}(-1,-1))={\cal O}(3,0)\otimes {\cal O}(0,3)\otimes {\cal O}(-1,-1)={\cal O}(2,2);\quad Ber_{\mathbb C}\Pi F={\cal K}_X\otimes det F={\cal O}$-is trivial. 

It implies that the bundle $Ber_{\mathbb C}\Pi F$ admits a nonvanishing section $vol_{\Pi F}$.

This section is $SO(4)$-invariant. The action of $\mathfrak{u}=\mathbb{C}+\mathbb{C}$-the unipotent subgroup of Borel subgroup $B\subset SO(4)$  on the large Schubert cell of $X$ is  transitive and free. 

Hence the section of $Ber_{\mathbb{C}} \Pi F$ in $\mathfrak{u}$-coordinates is 
$$vol_{\Pi F}=dz_l\wedge dz_r\wedge d\alpha _1\wedge d\alpha _2\wedge d\alpha _3\wedge d\tilde {\alpha} _1\wedge d\tilde {\alpha }_2\wedge d\tilde {\alpha} _3\wedge d\theta,$$ where $\alpha_1,\dots,\tilde {\alpha} _3,\theta$ are $\mathfrak{u}$-invariant coordinates on the odd fiber. The section $vol_{\Pi F}$ is in the kernel of  $D$ by construction.

We can construct on manifold  $\Pi F$  a global holomorphic integral $-2$-form the way explained in remark (\ref{E:hhghssa}).
In our case it is equal to $c_{\Pi F}=\alpha_1\dots\tilde{\alpha}_3\theta d\alpha_1\wedge \dots \wedge  d\tilde{\alpha}_3\wedge  d\theta$.

\begin{proposition}
The form $\mu=vol_{\Pi F} \otimes \bar {c}_{\Pi F}$ is $D$ closed nontrivial integral $(0,-2)$-form on the underlying real graded $\Pi F$.
\end{proposition} 
\begin{proof}
Direct inspection in local coordinates.

\end{proof}
An integral form $\mu$ defines a $\dbar+\alpha$-closed trace on $\Omega^{0\bullet}(\Pi F)\otimes Mat_n$ 
$$\int a=tr\int_{\Pi F}a\mu$$
\begin{definition}
By definition an A$\ity$ algebra algebra is a graded linear space , equipped with a series of maps $\mu_n:A^{\otimes n}\to A, n\ge 1$ of degree $2-n$ that  satisfy quadratic relation:

\begin{equation}
\begin{split}
&\sum_{i+j=n+1}\sum_{0\le l\le i}\epsilon(l,j)\times\\
&\mu_i(a_0,...,a_{l-1},\mu_j(a_l,...,a_{l+j-1}),a_{l+j},...,a_n)=0
\end{split}
\end{equation}
where $a_m\in A$, and

$\epsilon(l,j)=(-1)^{j\sum_{0\le s\le l-1}deg(a_s)+l(j-1)+j(i-1)}$.

In particular, $\mu_1^2=0$.

\end{definition}

\begin{remark}\label{E:gfdadfjj}
Suppose we have an A$\ity$ algebra $A$ equipped with a projector $\pi$.  A homotopy $H$ such that $\{d,H\}=id-\pi$ can be used as an input data for construction of  a new A$\ity$ structure on $\Imm \pi$ (see [Kad],[Markl] for details). The  homotopy $H$  is not unique. The resulting A$\ity$ algebras  will have different multiplications, depending on $H$. All of them  will be A$\ity$ equivalent. An additional structure on $A$  helps to fix an ambiguity in a  choice of $H$. In our case  algebra $A$ is a Dolbeault complex of a manifold with an operator $\pi$ being an orthogonal projection on cohomology. If the manifold is compact, K\"{a}hler and a $G$-homogeneous there is  natural choice of $H$: $H=\dbar^*/\Delta'$. The operator $\dbar^*$ $\Delta'$ are build by a $G$-invariant metric. The operator $\Delta'$ is equal to $\Delta$ on $\Ker \Delta^{\perp}$ and equal to identity on $\Ker \Delta$.
\end{remark}
\begin{remark}\label{E:gfdadfjjdsfd}
A construction  described in remark (\ref{E:gfdadfjj}) admits a generalization. Suppose an A$\ity$ algebra $A$ has a differential $d$ that is a sum of two anticommuting differentials $d_1$ and $d_2$. Assume that $\{d_1,H\}=id-\pi$ and  a composition $d_2H$ is a  nilpotent operator . Then $\Imm \pi$ carries a structure an A$\ity$ algebra quasiisomorphic to $A$. The same statement is true for A$\ity$ algebras with a trace.
The proof goes along the same lines as in ref. [Markl], but we allow two-valent vertices.
\end{remark}

Technically it is more convenient to work not with algebra $\Omega ^{0\bullet}(\widetilde{\Pi F})$ but with a quasiisomorphic subalgebra $(\Omega ^{0\bullet}\Lambda F,D)$ .

In application of the constructions from the remarks (\ref{E:gfdadfjj}, \ref{E:gfdadfjjdsfd}) we choose $\pi$ to be an orthogonal projector from the Hodge theory, corresponding to $SO(4,\mathbb{R})$-invariant metric on $X$. We also use  a decomposition $D=d_1+d_2=\dbar+\alpha$.

The algebra of cohomology of $(\Omega ^{0\bullet}\Lambda F,\dbar)$ carries an A$\ity$-algebra structure. We denote it by $C=H^{\bullet}(X,\Lambda(H^*)\otimes \Lambda(\Theta))$. Denote by $\psi$ a quasiisomorphism $(\Omega ^{0\bullet}\Lambda F,\dbar)\rightarrow C$. We shall describe some  properties of $C$. 
Let $W_l$, $W_r$ be spinor representations of $SO(4)$. The vector representation $V$ is equal to $W_l\otimes W_r$.

The differential $\alpha$  induces a differential $[\alpha]$ on $C$. The ghost grading of the group $H^i(X,\Lambda^k(H^*)\otimes \Lambda^s(\Theta))$ is equal to $i+s$, the additional grading is equal to $k+2s$(preserved by $\dbar$ and $\alpha$). We used a nonstandard ghost grading that differs from the one used in physics  on shift by one. In particular the ghost grading of the gauge (labeled by $V$) and spinor (labeled by spinors $W_l$, $W_r$) and matter ($SO(4)$ action is trivial ) fields is equal to one.   In the table below you will find  the field content (representation theoretic description ) of $C$ :
\begin{equation}\label{E:dkns}
\mbox{
\scriptsize{
$\begin{array}{lll|lll}
gh&deg&&gh&deg&\\
0&0&H^0(X,\Lambda^{0}(H^*))=\mathbb{C}&1&2&H^0(X,\Lambda^{0}(H^*)\otimes \Theta)=V\\
1&2&H^1(X,\Lambda^{2}(H^*))=\Lambda^2(T)+\Lambda^2(T^*)&1&3&H^0(X,\Lambda^{1}(H^*)\otimes \Theta)=W_l\otimes T+W_r\otimes T^*\\
1&3&H^1(X,\Lambda^{3}(H^*))=W_l+ W_r&1&4&H^0(X,\Lambda^{2}(H^*)\otimes \Theta)=T\otimes T^*\\
2&4&H^2(X,\Lambda^{4}(H^*))=T\otimes T^*&2&5&H^1(X,\Lambda^{3}(H^*)\otimes \Theta)=W_l+ W_r\\
2&5&H^2(X,\Lambda^{5}(H^*))=W_l\otimes\Lambda^2(T)+ W_r\otimes\Lambda^2(T^*)&2&6&H^1(X,\Lambda^{4}(H^*)\otimes \Theta)=T+T^*\\
2&6&H^2(X,\Lambda^{6}(H^*))=V&3&8&H^2(X,\Lambda^{6}(H^*)\otimes \Theta)=\mathbb{C}
\end{array}$
      }
      }
      \end{equation}
The groups $H^0(X,\Lambda^{2}(H^*)\otimes \Theta)$ and $H^2(X,\Lambda^{4}(H^*))$ are contracting pairs, they are killed by the differential $[\alpha]$ and should be considered as auxiliary fields in the related CS theory. 

An  A$\ity$ algebra $C$ besides differential $[\alpha]$ and  multiplication has higher multiplication on three arguments (corresponding to cubic nonlinearity of YM equation). However operations in more then three arguments are not present.
This can be deduced from homogeneity of $\dbar$ and $\alpha$ with respect to the additional grading. Finally representation theory fixes structure maps up to a finite number of parameters. The integral $\int$ defines a nonzero map $tr:H^2(X,\Lambda^{6}(H^*)\otimes \Theta)\rightarrow \mathbb{C}$.

Presumably it is possible to complete this line of arguments to a full description of multiplications in $C$. We prefer do it indirectly through the relation to Berkovits construction [Berk].

\begin{remark}
Let $\mathbb{R}^{10}$ be a linear space, equipped  with a positive-definite dot-product. Denote by $S$ an irreducible complex spinor representation of orthogonal group  $SO(10)$. Denote by $\Gamma_{\alpha\beta}^i$ coefficients of the nontrivial intertwiner $\Sym^2(S)\rightarrow \mathbb{C}^{10}$ in some basis of $S$ and an orthonormal basis of $\mathbb{C}^{10}$. 
We assume that  $\mathbb{C}^{10}$-basis is real.

On a superspace $(\mathbb{R}^{10}+ \Pi S)\otimes \mathfrak{u}(n)$ we  define a superfunction (Lagrangian)
\begin{equation}
S(A,\chi)=\sum_{i<j}tr([A_i,A_j][A_i,A_j])+\sum_{\alpha\beta i}tr(\Gamma_{\alpha\beta}^i[A_i,\chi^{\alpha}]\chi^{\beta})
\end{equation}
$A_1,\dots, A_{10}$ is a collection of antihermitian matrices labeled by the basis of $\mathbb{C}^{10}$. Similarly odd matrices $\chi^1,\dots,\chi^{16}$ are labeled by the basis of $S$.
This can be considered as a  field theory, obtained from D=10, N=1 YM theory by reduction to zero dimensions. We call it IKKT after the paper [IKKT] where it has been studied.  IKKT theory has a gauge invariance - invariance with respect to conjugation. A BV version  of IKKT coincides with a CS  theory associated with an A$\ity$ algebra $\mathcal{A}_{IKKT}$ that we shall introduce presently.
\end{remark}

\begin{definition}
An A$\ity$ algebra $\mathcal{A}_{IKKT}$ can be considered as vector space spanned
by symbols $x_{k}, \xi ^{\alpha }, c, x^{\ast k}, \xi _{\alpha }^{\ast }, c^{\ast }, 1\leq k\leq 10, 1\leq \alpha \leq 16$
with operations $\mu_{2}$ (multiplication), $\mu_{3}$(Massey product) defined by the following formulas:
\begin{align}
&\mu_2(\xi ^{\alpha }, \xi ^{\beta })=\Gamma^{\alpha\beta}_k x^{\ast k}\\
&\mu_2(\xi ^{\alpha }, x_{k})=\mu_2(x_{k}, \xi ^{\alpha })=
\Gamma^{\alpha\beta}_k\xi _{\beta }^{\ast }\\
&\mu_2(\xi ^{\alpha }, \xi _{\beta }^{\ast })=
\mu_2(\xi _{\beta }^{\ast }, \xi ^{\alpha })=c^{\ast }\\
&\mu_2(x_{k}, x^{\ast k})=\mu_2(x^{\ast k}, x_{k})=c^{\ast }\\
&\mu_3(x_{k}, x_{l}, x_{m})=\delta_{kl}x^{\ast m}-\delta_{km}x^{\ast l}\\
&\mu_2(c, \bullet)=\mu_2(\bullet, c)= \bullet \\
\end{align}
All other products are equal to zero. 
An element $c$ is a  unit.

All operations $\mu_{k}$ with $k\neq 2, 3$ vanish.
The algebra carries a trace functional $tr$ equal to one on $c^*$ and zero on the rest of the generators. It induces a dot-product by the formula $(a,b)=tr(\mu_2(a,b))$, compatible with $\mu_k$.
\end{definition}

By definition an A$\ity$ algebra has a grading (we call it a ghost grading ) such that  operation $\mu_n$ has degree $2-n$. An A$\ity$ algebra might also have an additional grading such that all operations have degree zero with respect to it. See [MSch05] for details on gradings of $\A_{IKKT}$.

\begin{proposition}\label{P:bbncnsq}

A differential graded  algebra with a trace $(\Omega ^{0\bullet}(\widetilde{\Pi F}),\dbar+\alpha,tr_{\mu})$
 is quasiisomorphic to $\mathcal{A}_{IKKT}$.
\end{proposition} 
\begin{proof}

We shall employ  methods  developed in [MSch05]. Recall that the manifold of pure spinors in dimension ten is equal to $\Q=SO(10,\mathbb{R})/U(5)$. As a complex manifold it is defined as the space of solutions of homogeneous equations $\Gamma_{\alpha\beta}^i\lambda^{\alpha}\lambda^{\beta}=0$, where $\lambda^{\alpha}$ are homogeneous coordinates on $\mathbf{P}^{15}$. A space $\mathbf{P}^{15}$ is the projectivisation of irreducible complex spinor representation of $Spin(10,\mathbb{R})$.  
We denote by $R$ the  restriction on $\Q$ of the twisted tangent bundle $T_{\mathbf{P}^{15}}(-1)$.  Denote by $A$ the coordinate algebra $\mathbb{C}[\lambda^{1},\dots,\lambda^{16}]/\Gamma_{\alpha\beta}^i\lambda^{\alpha}\lambda^{\beta}$ of $\Q$. Denote by $B$ the  Koszul complex $A\otimes \Lambda[\theta^{1},\dots\theta^{16}]$ with differential $\lambda^{\alpha}\pr{\theta^{\alpha}}$.  

The algebra $B$ can be equipped with various gradings. The cohomological grading is defined on generators as $|\lambda^{\alpha}|=2$, $|\theta^{\alpha}|=1$; the grading by the degree (or homogeneous grading) is $deg\lambda^{\alpha}=deg\theta^{\alpha}=1$. The differential has degree one with respect to $||$ and zero with respect to $deg$. As a result the cohomology groups are bigraded: $H(B)=H^{ij}$, where $i$ corresponds to $||$-grading, $j$ corresponds to $deg$.

In [MSch05] we proved  that  $\bigoplus_{i-j=k} H^{ij}(B)=H^k(\Omega^{0\bullet}(\Q)\Lambda(R^*))$. In fact we proved that the identification  map is a quasiisomorphism of differential graded algebras with a trace.
In the language of supermathematics we may say that cohomology $H^k(\Omega^{0\bullet}(\Q)\Lambda(R^*))$ is Dolbeault cohomology of a split supermanifold $\Pi R$. 

In the course of the proof of quasiisomorphism we have identified the algebra of functions on the  fiber of projection 
\begin{equation}
\Pi R  \rightarrow \Q
\end{equation}
 over a point $pt=(\lambda_0^{\alpha})\in \Q$   with cohomology of algebra $B^{\bullet}_{pt}=(\Lambda[\theta^{1},\dots\theta^{16}],d)$ where $d=\lambda_0^{\alpha}\pr{\theta^{\alpha}}$. An  analog of the complex  $B^{\bullet}_{pt}$ can be defined for any subscheme $U$ of $\Q$ as a tensor product $B^{\bullet}_{U}=A_{U}\otimes \Lambda[\theta^{1},\dots\theta^{16}]$. The algebra $A_U$ is equal to $\bigoplus_{i\geq 0} H^0(U,\O(i))$. The cohomology of $B_U$ is equal to $ H^{\bullet}B_{pt}\otimes \O(U)$ if $\O(1)$ is trivial on $U$ .

We proved in  [MSch06] that the algebra $B$ with Berkovits trace $tr$ is quasiisomorphic to algebra $\A_{IKKT}$.  

We need to present a  useful observation from [MSch05]. Let us decompose a set $\{\lambda^{1},\dots,\lambda^{16}\}$ into a union $\{\lambda^{\alpha_1},\dots,\lambda^{\alpha_s}\}\cup\{\lambda^{\beta_1},\dots,\lambda^{\beta_k}\}$ such that $\{\lambda^{\alpha_1},\dots,\lambda^{\alpha_s}\}$ is a regular sequence. Then the algebras $(B,d)$ and $(B',d)=(A/(\lambda^{\alpha_1},\dots,\lambda^{\alpha_s})\otimes \Lambda[\theta^{\beta_1},\dots,\theta^{\beta_k}],d)$ are quasiisomorphic.

 The following construction has been described in [MSch05].The spin representation $S$ of $\mathfrak{so}(10)$ splits after restriction on $\mathfrak{gl}(3)\times \mathfrak{sl}_l(2)\times \mathfrak{sl}_r(2)$ into $T\otimes W_l+T^*\otimes W_r+W_l+W_r$. We choose  coordinates on $W_l+W_r$- equal to  $(\lambda^{\alpha_i})=(\tilde{w}_l^+,\tilde{w}_l^-,\tilde{w}_r^+,\tilde{w}_r^-)$. They form  a regular subsequence of  $\lambda^{\alpha}$ . The  manifold  corresponding to $A/(\lambda^{\alpha_i})$ is equal to $Q\cap \mathbf{P}(T\otimes W_l+T^*\otimes W_r)$. The  intersection is isomorphic to $F(1,2)\times X$. The algebra of homogeneous functions $A'=A(F(1,2)\times X)$ on $F(1,2)\times X$ is generated by $s^{i\alpha},t^j_{\alpha }(1\leq \alpha\beta \leq 3, 1\leq ij\leq 2)$. The relations are 
\begin{equation}
\sum_{\alpha}s^{i\alpha}t^j_{\alpha}=0, det(s^{i\alpha})=0, det(t^j_{\alpha})=0
\end{equation}
 In the formula $det$ stands for a row of $2\times 2$ minors of $2\times 3$ matrix.

 We plan to follow almost the same method of construction of supermanifold as for the Koszul complex of pure spinors. Denote by $g$ a projection $$g:F(1,2)\times X\rightarrow X$$ We fix a point $x\in X$. The algebra $A'_{g^{-1}(x)}$ is isomorphic to \\$\mathbb{C}[p_1,\dots p_3,u^1\dots,u^3]/(p_iu^i)$. The algebra $B'_{p^{-1}(x)}$ is isomorphic to 
\begin{equation}
A'_{g^{-1}(x)}\otimes\Lambda[\pi_1,\dots,\pi_3,\nu^1,\dots,\nu^3,\tilde{\pi}_1,\dots,\tilde{\pi}_3,\tilde{\nu}^1,\dots,\tilde{\nu}^3],d
\end{equation}
with a differential 
\begin{equation}\label{E:asdfhhjdjd}
d=p_i\pr{\tilde{\pi}_i}+u^i\pr{\tilde{\nu}^i}
\end{equation}
The cohomology of this differential  is equal to $$\Lambda[E_x]=\Lambda[\pi_1,\dots,\pi_3,\nu^1,\dots,\nu^3,\theta]$$ The induced A$\ity$ algebra structure on cohomology  has no higher multiplications.
The element $\theta$ is represented by a cocycle $\tilde{\pi_i}u^i$. The linear space  $E_x$ coincides with the fiber $F_x$ of vector bundle $F$ (\ref{E:sjs}). 
The main distinction between this computation and a computation with pure spinors  is that we encountered  a noncanonical A$\ity$ morphism  $\iota:B'_{p^{-1}(x)}\rightarrow \Lambda[E_x]$, which could be  not a homomorphism of associative algebras.

Recall that we viewed the cohomology of $B_{pt}$ as functions of the fiber of projections $p:\Pi R \rightarrow \Q$. We have a natural identification of fibers over different patches of $\Q$. It gives us a consistent  construction of a  split manifold $\Pi R$.   

In case of the manifold $X$ if we ignore the issues related to ambiguities  of  choice of morphism $\iota$ tt is not hard to see that an  isomorphism of fibers  $F_x\cong H_x$ can be extended to a $Spin(4)$ equivariant isomorphism of the  vector bundles $F_x\cong H_x$ (use homogeneity of both vector bundles with respect to  $\mathfrak{sl}^l(2)\times \mathfrak{sl}^r(2)$ action).

 This way we recover the manifold $\Pi F$. In reality when we try to glue rings of functions on different patches the structure isomorphisms will be A$\ity$-morphisms. We may claim on general grounds  that we  get an A$\ity$ structure on a space of \v{C}ech chains of $\Pi H\cong \Pi F$. This structure can be trivialized  by a twist on a local A$\ity$-morphism (reduced to the standard multiplication in Grassmann algebra) on every double  intersection $U_{ij}$ of patches (if $U_{ij}\subset X$ is sufficiently small). An ambiguity in a choice of such twist leads to appearance of \v{C}ech 2-cocycle $\beta_{ijk}$ with values in infinitesimal (not A$\ity$ ) transformations of the fiber $\Pi F$.
In the Dolbeault picture this cocycle corresponds to $\alpha$. Finally we use \v{C}ech-Dolbeault equivalence. This proves the claim.

\end{proof}

\begin{remark}
The algebra $C$ carries a differential $d=[\alpha]$. The minimal model of $C$ (by definition it is a quasiisomorphic A$\ity$ algebra without a differential) constructed for an obvious homotopy of differential $d=[\alpha]$ is quasiisomorphic to $\A_{IKKT}$. If we ask for a quasiisomorphism to be compatible with all gradings that exist on both algebras this quasiisomorphism is an  isomorphism.

From this is is quite easy to recover all multiplications in the algebra $C$.
\end{remark}

\section{Nonreduced theory}\label{S:ffkewx}
A manifold $Z$ with an integral form is constructed in this section.

\subsection{Construction of the algebra $A(Z)$}\label{S:wfdke}
In this section we construct an algebra $A(Z)$. It will be a linking chain between YM theory and ambitwistors.

We construct a manifold $Z$ as a direct product $\Pi F\times \Sigma$. Intuitively speaking the algebra $\Omega^{0\bullet}(\widetilde{\Pi F})$ carries all information about YM theory reduced to a point, whereas  algebra $C^{\infty}(\Sigma)$ contains similar information about the space $\Sigma$. The idea is that  a tensor product $\Omega^{0\bullet}(\widetilde{\Pi F})\otimes C^{\infty}(\Sigma)$ with a suitably twisted differential will contain all  information about 4-D YM theory.
Later we will interpret the same complex as tangential CR complex on the manifold $Z$.

The linear space 
\begin{equation}\label{E:gdfhd}
\Sigma^{\mathbb{C}}=V=W_l\otimes W_r
\end{equation}
 has coordinates $ x^{ij}, 1\leq i,j \leq 2$. The vector space $V$ has an $SO(4,\mathbb{C})$  action compatible with decomposition (\ref{E:gdfhd}). It is  induced from the $SO(4,\mathbb{R})$ action on $\Sigma$ .

Define  a  differentiation of $\Omega ^{0\bullet}(\widetilde{\Pi F})\otimes C^{\infty}(\Sigma)$ as follows.

There is an $SO(4)$ equivariant isomorphism $V\cong H^0({\cal O}(1,1))$.
The line bundle ${\cal O}(1,1)$ is generated by its global sections.
We have a short exact sequence:
\begin{equation}\label{Q:wert}
0\rightarrow M \rightarrow V\overset{m}{\rightarrow}\Theta={\cal O}(1,1)\rightarrow 0
\end{equation}
where $V$ is considered as a trivial vector bundle with fiber $V$.  The differentiation $\delta$ of $\Omega ^{0\bullet}(\widetilde{\Pi F}) \otimes C^{\infty}(\Sigma)$ is equal  $\delta=m(x^{ij})\pr{x^{ij}}$. We interpret $\pr{x^{ij}}$ as global sections of $T^{\mathbb{C}}\Sigma$. 

Recall that the  differential  $D$ in $\Omega ^{0\bullet}(\widetilde{\Pi F})$ is equal to $\dbar+\alpha$.
It becomes clear from explicit computation of cohomology that coefficients $\beta^{ij}$ in $\{\alpha, \delta\}=\beta^{ij}\pr{x^{ij}}$ are $\dbar$-exact.

 Choose $\gamma^{ij}$ such that $\dbar \gamma^{ij}=-\beta^{ij}$. Define a  differentiation $\gamma$ of $\Omega ^{0\bullet}(\widetilde{\Pi F})\otimes C^{\infty}(\Sigma)$ as $\gamma^{ij}\pr{x^{ij}}$ and zero on the rest of the generators. 

It follows from our construction that $D_{ext}=D+\delta+\gamma$ satisfies $D_{ext}^2=0$ . Denote $D'=\delta+\gamma$.

By definition the integral form on manifold $\widetilde{\Pi F}\times \Sigma$ is :
\begin{equation}
Vol=\mu\otimes \bigotimes^2_{i,j=1} dx^{ij}
\end{equation}

\begin{proposition}
$Vol$ is invariant with respect to $D$, $D'$ and therefore with respect to $D_{ext}$.
\end{proposition}
\begin{proof}
Direct inspection.
\end{proof}
\subsection{Proof of the equivalence} 

\begin{proposition}\label{P:jfjhsg}

Denote $Z=\Pi F\times \Sigma$. We equip algebra  $A(Z)=\Omega ^{0\bullet}(\widetilde{\Pi F})\otimes C^{\infty}(\Sigma)\otimes Mat_n$ with a trace
\begin{equation}
\int a=tr\int_{Z}a Vol
\end{equation}

The CS theory constructed by the triple $(A(Z),D_{ext},\int)$ is classically equivalent to N=3 euclidean YM  with a gauge group U$(n)$.
\end{proposition}
 The equivalence should hold also on a quantum level.
\begin{proof}
We shall only outline the basic ideas.

A precise mathematical statement is about quasiisomorphism of certain A$\ity$ algebras. One of them is $A(Z)$.  The reader should consult [MSch06] for information about A$\ity$ algebra with a trace corresponding to N=3 YM theory.

Suppose $\psi:A\rightarrow B$ is a quasiisomorphism of two A$\ity$ algebras. Let $m$ be an associative algebra. Then we have a quasiisomorphism of tensor products $\psi:A\otimes m\rightarrow B\otimes m$. Let $a$ be a solution of MC equation in  $A\otimes m$ (the reader may consult [MSch05] for the definition). The map $\psi$ transports  solution $a$ to a solution $\psi(a)$ of MC equation in $B\otimes m$.

In a formal interpretation of structure maps of A$\ity$ algebra $A$ as the Taylor coefficients of a noncommutative vector field on noncommutative space $\mathbb{A}$ solution $a$ of MC equation corresponds to a zero of the vector field. We  can expand the  vector field into series at  $a$ and get some new A$\ity$ algebra. This construction is particularly transparent in case of a dga. A solution of MC equation defines a new differential $\tilde{d}x=dx+[a,x]$. It corresponds to a shift of a vacuum in a physics jargon. Denote by $A_a$ an A$\ity$ algebra constructed by the element $a$ .

It is easy to see that the map $\psi$ defines (under some mild assumptions in $a$) a quasiisomorphism $$\psi:A\otimes m_a\rightarrow B\otimes m_{\psi(a)}$$

Denote by $Diff(\Sigma)$ an algebra of differential operators on $\Sigma$.
We would like to apply  construction from the previous paragraph to the tensor products $\Omega^{0\bullet}(\widetilde{\Pi F})\otimes Diff(\Sigma)$ and $C\otimes Diff(\Sigma)$. We can interpret $\delta+\gamma$ as a solution of MC equation for algebra $\Omega^{0\bullet}(\widetilde{\Pi F})$ with coefficients in $Diff(\Sigma)$. The  quasiisomorphism $\psi$ maps $\delta+\gamma$ into an element $y^{ij}\pr{x^{ij}}$, where $y^{ij}$ is a basis of $H^0(X,\Lambda^{0}(H^*)\otimes \Theta)\subset C$.

The rest is a  matter of formal manipulations. It is straightforward to see that $C\otimes Diff(\Sigma)_{\psi(a)}$ is an A$\ity$ mathematics counterpart of YM equation where the gauge potential, spinors, matter fields are having  their coefficients not in functions on $\Sigma$ but in $Diff(\Sigma)$.
This is not precisely what we have hoped to obtain.  We shall address this issue presently.

Suppose an associative algebra $m$ contains a subalgebra $m'$. The previous construction has a refinement. The noncommutative vector field on a space $\mathbb{A}_m$ corresponding to A$\ity$ algebra $A\otimes m$ is tangential to a noncommutative subspace $\mathbb{A}_{m'}$ (because $m'$ is closed under multiplication). We say that a solution of MC $a\in A\otimes m$ is compatible with $m'$ if the vector field defined by the algebra $ A\otimes m$ is tangential to the space $a+\mathbb{A}_{m'} \subset \mathbb{A}_m$. This is merely  another way to say that a linear space $A\otimes m'$ is a subalgebra of $(A\otimes m)_a$. If $a$ is compatible with $m'$ the map  $\psi$ (under some mild assumptions on $a$) induces a quasiisomorphism $\psi:A\otimes m'\rightarrow B\otimes m'_{\psi(a)}$. 

We apply this construction to subalgebra $C^{\infty}(\Sigma)\subset Diff(\Sigma)$ for which mentioned above condition on   $\delta+\gamma$  is met. 

Suppose in addition that algebras $A,B,m'$ have a traces and a morphism $\psi:A\rightarrow B$ is compatible with the traces. Assume moreover that the induced A$\ity$ structure $A\otimes m'_{a}$ is compatible with a trace.  Then the induced morphism $\psi:A\otimes m'_a\rightarrow B\otimes m'_{\psi(a)}$ is compatible with traces.

In our case the operator $D'$ preserves the integral form $Vol$  and the above conditions are met.

The CS theory associated with the algebra $C\otimes C^{\infty}(\Sigma)$ has  the following even part of the Lagrangian
$$<F_{ij},F_{ij}>+<\nabla_i\phi^{\alpha},\nabla_i\phi_{\alpha}>+<[\phi^{\alpha},\phi_{\beta}],K_{\alpha}^{\beta}>+<K_{\alpha}^{\beta},K^{\alpha}_{\beta}>$$
The theory besides of the gauge field corresponding to connection $\nabla_i$ in a principle U$(n)$ bundle , matter fields $\phi^{\alpha}\in  Ad\otimes T,\phi^{\beta}\in Ad\otimes T^*$ contains an auxiliary field $K_{\alpha}^{\beta}\in Ad\otimes T\otimes T^*$. This theory is equivalent to N=3 YM( the odd parts of the Lagrangians coincide ).

\end{proof}

\subsection{Relation between a CR structure on $Z$ and an algebra $A(Z)$}\label{S:gsfre}

In this section we give a geometric interpretation of the algebra $A=\Omega^{0\bullet}(\widetilde{\Pi F})\otimes C^{\infty}(\Sigma)$.

Fibers of projection 
\begin{equation}\label{E:cjsgwu}
p:X\times \Sigma\rightarrow \Sigma
\end{equation}
have holomorphic structure. Denote $T_{vert}$ a bundle of $p$-vertical vector fields. We  define a distribution $\G$ as $T_{vert}^{1,0}\subset T_{vert}^{\mathbb{C}}$

Choose a linear basis $e_1,\dots,e_4\in \Sigma$. Define $\pr{x^s}$ the differentiations in the direction of $e_s$.
Restrict a map $m$ from short exact sequence (\ref{Q:wert}) on $\Sigma \subset V$, then $m(e_s)$ is a set of holomorphic sections of $\Theta$.

For any point $x\in X$ we have a subspace ${\cal H}_x\subset T^{\mathbb{C}}_{\Sigma}$ spanned by 
\begin{equation}
\sum_{i=1}^4m(e_s)_x\pr{x^d}
\end{equation}
A union of such subspaces  defines  a complex distribution ${\cal H}$ on $X\times \Sigma$ .

Define an integrable distribution $\F={\cal G}+{\cal H}\subset T^{\mathbb{C}}_{X\times \Sigma}$.

The reader can see that the CR structure on the space $X\times \Sigma$ literally coincides with the CR structure on ambitwistor space $A(\Sigma)$ defined in Section \ref{S:hfdgjpo}.

In light of this identification an  element $\theta$ (a local coordinate on $\Theta^*$) can be interpreted as a local CR-form with nonzero values on $\overline{{\cal H}}$.

Restriction  of  the vector bundle $H^*$ (used in the construction of supermanifold $F$ in equation (\ref{E:sjs})) on $X\times \Sigma$ is holomorphic along $\F$. Additionally we can interpret component $\alpha$ in the differential $D$ on $\Omega ^{0\bullet}(\widetilde{\Pi F})$ as a contribution from a  superextension of the CR structure defined in (\ref{E:gfdjfur}).

From this we deduce that the algebra $A(Z)=(\Omega ^{0\bullet}(\widetilde{\Pi F})\otimes C^{\infty}(\Sigma),D_{ext})$ we have constructed coincides with the CR tangential complex on an open subset of the manifold  $\RH $.

The integral form $Vol$ is the only CR-holomorphic form invariant with respect to $SU(2)\times SU(2)\ltimes \mathbb{R}^4$.

 From this we deduce that it coincides (up to a multiplicative constant) with the integral form defined by $vol$

It is possible reconstruct an action of the super Poincare group SP on $A$. The reader will find  explanations of why the measure is not invariant with respect to the full superconformal group in Section \ref{S:gufjasklly}.

\section{Appendix}

\subsection{On the definition of a graded real superspace}\label{S:llfddfgg}
A tensor category of complex superspaces $\CC$ (see ref.[DMiln] for introduction to tensor categories) has two real forms. The first is a category of real superspaces $\CR$. It is more convenient to think about objects of this category as of  complex superspaces, equipped with an antiholomorphic involution $\sigma$.

Another tensor category related to $\CC$ is formed by complex superspaces, equipped with an antilinear map $\rho$
\begin{definition}\label{D:gdfjgdf}
Suppose $V$ is a $\mathbb{Z}_2$-graded vector space over complex numbers. An antilinear map $\rho:v\rightarrow \bar{v}$ is  a graded real structure if 
\begin{equation}
\begin{split}
&\rho^{2}=sid\\
& sid(v)=(-1)^{|v|}v
\end{split}
\end{equation}
we  denote by $|v|$ a parity of $v$.

An element $v$ is real iff $\bar v=v$. Only even elements can be real with respect to a graded real structure. A graded real superspace is a pair $(V,\rho)$. Grader real superspaces form a tensor category $\CH$

\end{definition}

 We shall be mostly interested in   categories $\CC$ and $\CH$.

The categories $\CC$ $\CH$ are related by tensor functors. 

The first functor is   complexification $\CH\Rightarrow \CC $, $V\Rightarrow V^{\mathbb{C}}$. It  forgets about the map $\rho$.

The second  functor is $\CC \Rightarrow \CH$, $V\Rightarrow V^{\mathbb{H}}$. The object $V^{\mathbb{H}}$ is a direct sum $V+\overline{V}$. There is an antilinear isomorphism $\sigma:V\rightarrow \overline{V}$. For $v=a+b+c+d\in V^0+V^1+\overline{V}^0+\overline{V}^1$ define 
\begin{equation}\label{E:qdsfdfbshe}
\rho(a+b+c+d)=\sigma^{-1}(c)-\sigma^{-1}(d)+\sigma(a)+\sigma(b)
\end{equation}
By construction $\rho^2=sid$.

A language of tensor categories can be used as a foundation for developing  Commutative Algebra and Algebraic Geometry. If we start off in this direction with a category $\CC$ the result will turn to be  Algebraic Supergeometry.  

A category $\CH$  provides us with some real form of this geometry. 

First of all a $\CH$ or a real graded  manifold is  an algebraic supermanifold $M$ defined over  $\mathbb{C}$. The manifold $M$ carries some additional structure. The manifold $M_{red}$ is equipped with an antiholomorphic involution $\rho_{red}$. There is also an antilinear isomorphism of sheaves of algebras  
\begin{equation}\label{E:jjshdga}
\rho:\rho_{red}^*\O\rightarrow \O,
\end{equation}
 such that $\rho^2=sid$.

 There is a $C^{\infty}$ version of a $\CH$ manifold. It basically mimics a stricture of $C^{\infty}$ completion of algebraic $\CH$ manifold at locus of  $\rho_{red}$ fixed points. Any $C^{\infty}$ manifold admits a noncanonical splitting: $M\cong \Pi E$, where $E$ is some complex vector bundle over $M_{red}$. A $\CH$-structure manifests in an antilinear automorphism $\rho$ of $E$, that satisfies $\rho^2=-id$. Observe that $\rho$, together with multiplication on $i$ defines a quaternionic structure on $E$.

For any complex algebraic supermanifold $M$ there is the underlying  $\CH$ manifold. As an algebraic supermanifold it is equal to $M\times \overline M$. There is a canonical antiinvolution on  $(M\times \overline M)_{red}$. The morphism of sheaves (\ref{E:jjshdga}) in local charts is defined by formulas similar to (\ref{E:qdsfdfbshe}).

This construction manifests itself in $C^{\infty}$-setting as follows. Any complex supermanifold $M$ defines  a $C^{\infty}$ supermanifold $\tilde M$ that is holomorphic in odd directions. It is a $C^{\infty}$-completion  of $M\times \overline{M}_{red}$  near diagonal of $M_{red}\times \overline{M}_{red}$.  Suppose $\Pi E$ is a splitting of  $\tilde M$, where $E$ is a complex vector bundle on $\tilde{M}_{red}$. The vector bundle $E+\overline{E}$ has a natural quaternionic structure and defines a $C^{\infty}$ $\CH$-manifold $\Pi(E+\overline{E})$. This manifold is isomorphic to completion of $M\times \overline{M}$. Sometimes it is more convenient to work with the manifold $\tilde{M}$.

\subsection{On homogeneous CR-structures}\label{S:gfgjheq}
Suppose we are given an ordinary real Lie group and a closed subgroup $A\subset G$ with Lie algebras $\mathfrak{a}\subset \mathfrak{g}$. Additionally we have a complex subgroup $P\subset G^{\mathbb{C}}$ in complexification of $G$, with  Lie algebras $\mathfrak{p}\subset\mathfrak{g}^{\mathbb{C}}$. We assume  that the map 
\begin{equation}\label{E:frtw}
p:G/A\rightarrow G^{\mathbb{C}}/P
\end{equation} 
is a local  embedding. By construction $G^{\mathbb{C}}/P$ is a holomorphic homogeneous space. It  tangent space $T_x, x\in G/A$ contain a subspace $H_x=T_x\cap JT_x$. The operator $J$ is an operator of complex structure on $G^{\mathbb{C}}/P$. Due to $G$-homogeneity  spaces $H_x$ have constant rank and form a subbundle $H\subset T$.  We can decompose $H\otimes \mathbb{C}=\F+\overline{\F}$. It follows from the fact that $\mathfrak{p}$ is a subalgebra that the constructed distribution $\F$ is integrable and defines a CR-structure. 

A condition that the map $p$ (\ref{E:frtw}) is a local embedding is equivalent to $\mathfrak{g}\cap \mathfrak{p}=\mathfrak{a}$. In other words
\begin{equation}
\mathfrak{p}\cap\bar{\mathfrak{p}}=\mathfrak{a}^{\mathbb{C}}
\end{equation} 
It is easy to see that a fiber $\F_x$ at a point $x$ is isomorphic to $\mathfrak{p}/\mathfrak{a}^{\mathbb{C}}$

This construction of CR-structure  can be extended to a  supercase. A consistent way to derive  such extension   is to use a  functorial language of ref. [DM].

 However since this exercise, which we leave to the interested reader, is quite straightforward we provide only the upshot.  

We start off with description of  a data that  defines  a homogeneous space of a supergroup.

A complex homogeneous space $X$ of a complex supergroup $G$ with Lie algebra $\mathfrak{g}$ is encoded by :

{\bf 1.Global data:} An isotropy subgroup  $H\subset G_{red}$ (closed, analytic, possibly nonconnected ). This data defines  an ordinary  homogeneous space $X_{red}=G_{red}/H$;

{\bf 2.Local data:} a pair of complex super Lie algebras $\mathfrak{p}\subset \mathfrak{g}$ such that $\mathfrak{p}_0=Lie(H), Lie(G_{red})=\mathfrak{g}_0$ .

In the cases when we specify only Lie algebra of isotropy subgroup is clear from the context. 

 A real  graded structure on a  homogeneous space $X=G/A$ is encoded by  an antiholomorphic involution  on $G_{red}$ that leaves subgroup $A_{red}$ invariant;  a graded real structure $\rho$ on $\mathfrak{g}$,such that $\rho(\mathfrak{a})\subset\mathfrak{a}$.

If we are given a real subalgebra $(\mathfrak{a},\rho)\subset (\mathfrak{g},\rho)$ and a complex subalgebra $\mathfrak{p}\subset\mathfrak{g}$ such that $a=\mathfrak{p}\cap \rho(\mathfrak{p})$ we  claim that a  supermanifold $G/A$ carries a $(G,\rho)$-homogeneous CR-structure.
\subsection{General facts about CR-structures on supermanifolds}\label{S:gufjasklly}
In this section we will discuss mostly general facts about CR structures specific to supergeometry.
Suppose $M$ is a supermanifold equipped with an integrable CR distribution $\F$.
We present some basic examples of $\F$-holomorphic vector bundles on $M$.

{\bf Example} Sections of vector bundle $T^{\mathbb{C}}/\overline{\F}$ is a module over Lie algebra of sections of $\overline{\F}$. Thus the gluing cocycle of this bundle is CR-holomorphic. It implies that the bundle $\Ber$ is also CR-holomorphic.

Suppose we have a trivial CR-structure on $\mathbb{R}^{n_1|n_2}\times\mathbb{C}^{m_1|m_2}$. We assume that the space is equipped with global coordinates $x_i,\eta_j,z_k,\theta_l$.   The algebra of tangential CR complex $\Omega_{\F}^{\bullet}(\mathbb{R}^{n_1|n_2}\times\mathbb{C}^{m_1|m_2})$ has topological generators $x_i,\eta_j,z_k,\theta_l,\bar{z}_k,\bar{\theta}_l,d\bar{z}_k,d\bar{\theta}_l$. Denote by $A$ a subalgebra generated by $x_i,\eta_j,z_k,\theta_l,\bar{z}_k,,d\bar{z}_k$ and by $K$ a subalgebra generated by $\bar{\theta_l},d\bar{\theta}_l$. We have $\Omega^{0\bullet}=A\otimes K$. The algebra $K$ has trivial cohomology. As a result   projection $\Omega^{0\bullet}\rightarrow A$ is a quasiisomorphism.

It turns out that this construction exists in a more general context of an arbitrary CR-manifold. Informally we may say  think that a super-CR manifold is affine in holomorphic odd directions. It parallels with the complex case.

The construction requires a choice of $C^{\infty}$-splitting of CR-manifold $M$.

Suppose $Y$ is an ordinary manifold, $E$ is a vector bundle. Let $\Pi E$ denote the supermanifold whose sheaf of functions coincides with a sheaf of sections of the Grassmann algebra of the bundle $E^*$.  Such supermanifold is called split. By  construction it admits projection $p:\Pi E\rightarrow Y$. Any $C^{\infty}$ manifold is split, but the splitting is not unique. A space of global functions on $\Pi E$ is isomorphic to a space of sections of the Grassmann algebra  $\Lambda E^*$ of vector bundle $E$.

To make a connection with our considerations we identify  $Y=M_{red}$. 

Denote $\barF_{red}=\barF_{red}^0+\barF_{red}^1$ restriction of $\barF$ on $M_{red}$.  In terms of the splitting  operator $\dbar$ can be encoded by  a pair of operators of the first order $\dbar_{ev}:C^{\infty}(M_{red})\rightarrow \Lambda E^*\otimes \barF_{red}, \dbar_{odd}:E^*\rightarrow \Lambda E^*\otimes \barF_{red}$.

The lowest degree  component in powers $\Lambda^i E^*$ of the operator $\dbar_{odd}$ is  $\dbar^0_{odd} E^*\rightarrow \barF_{red}^1$.  It is a $C^{\infty}(Y)$ linear map, with a  locally free image. The image $S_0$ of a splitting $\barF^1_{red}\rightarrow  E^*$ can be used to generate a differential ideal $(S)$ of $\Omega_{\F}$. Denote the quotient $\Omega_{\F}/S$ by $\Omega_{s\F}$.

The complex $\Omega_{s\F}$ is a differential graded algebra. We can interpret it as a space of functions on some superspace $L$. A possibility to split $\dbar^0_{odd}$ implies smoothness of $L$.
\begin{proposition}\label{E:fdsafjh}
The map $\Omega_{\F}\rightarrow \Omega_{s\F}$ is a quasiisomorphism.
\end{proposition}
\begin{proof}
Follows from consideration of a spectral sequence associated with filtration $F^i\Omega^p_{\F}=(S_0)^{i-p}\Omega^p_{\F}$(we denote by $(S_0)^k$ the k-th power of the ideal generated by $S_0$ ).
\end{proof}
A CR-manifold $M$ is locally embeddable to $\mathbb{C}^{m|n}$ if in a neighborhood of a point there is a collection of $z_1,\dots,z_m$ even and   $\theta_1,\dots,\theta_n$ odd function that  are annihilated by $\dbar$ and whose Jacobian is  nondegenerate. 
\begin{definition}\label{D:wpprnfj}
Let us assume that  $\F^1|_{M_{red}}+\barF^1|_{M_{red}}=T^{\mathbb{C}1}$(superscript $1$ denotes the odd part),i.e. dimension of the odd part of CR distribution is maximal possible. 

We can locally generate ideal $S_0$ by elements $\overline{\theta}_1,\dots,\overline{\theta}_n$ and take $S$ as a $\dbar$ closure of $S_0$. It is not hard to check that under such assumptions Proposition \ref{E:fdsafjh} holds. Denote by $sM$ a submanifold specified by $S_0$
\end{definition}

\begin{remark}\label{E:fdfdsdfv}
 The Lie algebra of infinitesimal automorphisms of CR-structure is equal to $Aut_{\F}=\{a\in T^{\mathbb{C}}|[a,b]\in \barF, \mbox{ for all }b\in \barF\}\}$ with $Out_{\F}=Aut_{\F}/\barF$. In purely complex case the  quotient construction can be replaced by  $Out_{complex}=\{a\in \F|[a,b]\in \barF, \mbox{ for all }b\in \barF\}\}$ and the extension 
\begin{equation}
0\rightarrow \Inn_{\F} \rightarrow Aut_{\F}\rightarrow  Out_{\F}\rightarrow 0
\end{equation} has a splitting. By construction elements $\overline{\theta}_1,\dots,\overline{\theta}_n$ are invariant along vector fields from  distribution $\F$. We can guarantee that differential ideal generated by $\overline{\theta}_i$ is invariant with respect to elements of $Out_{complex}$. As a result we can push the action of $Out_{complex}$ to $\Omega^{\bullet}_{s\F}$-this is familiar fact from super complex geometry .  This contrasts with absence  of an action of $Aut_{\F}$ or $Out_{\F}$ on the ideal $S$ and on  $\Omega^{\bullet}_{s\F}$ for a general CR structure. One can prove however that  $\Omega^{\bullet}_{s\F}$ admits  an A$\ity$ action of $Aut_{\F}$ . A partial remedy is to consider subalgebra $\widetilde{Out}_{\F}=\{a\in \F|[a,b]\in \barF, \mbox{ for all }b\in \barF\}\}\subset Out_{\F}$. This subalgebra acts upon $\Omega^{\bullet}_{s\F}$ . However  this algebra is trivial if the Levi form of $\F$ is not degenerate
\end{remark}

Supergeometry provides us with a complex of CR-integral forms. Let $\Lambda \barF$ be the super Grassmann algebra of $\barF$. Let $Ber$ be the Berezinian  line bundle of a real manifold  $M$. Denote $\ ^{int}\Omega^{-p}_{\F}$ the tensor product $Ber\otimes \Lambda \barF$. There is a  pairing $(\omega,\nu)=\int_{M}<\omega,\nu>$ between sections $\omega\in \Omega^{p}_{\F}$ and $\nu\in \ ^{int}\Omega^{-p}_{\F}$. The symbol $<.,.>$ denotes contraction of a differential form with a  polyvector field. The operation  $<.,.>$ takes values in sections of $Ber$. The value $<\omega,\nu>$ can be used as an integrand for integration over $M$. 

The  orthogonal complement to the ideal $S\subset \Omega^{p}_{\F}$ is a subcomplex $\ ^{int}\Omega^{-p}_{s\F}\subset\ ^{int}\Omega^{-p}_{\F}$. It is  a differential graded module over $\Omega^{p}_{s\F}$. 
\begin{proposition}\label{P:fdfdhw}
Let $M$ be a supermanifold with a CR distribution $\F$ of dimension $(n|k)$. 
There is an isomorphism $i: \ ^{int}\Omega^{p-n}_{s\F} \rightarrow \Omega^{p}_{s\F}\Ber$ compatible with a stricture of  $\Omega^{\bullet}_{s\F}$-module.
The isomorphism is unique.
\end{proposition}
\begin{proof}

Using $C^{\infty}$ splitting we can identify $\Omega^{\bullet}_{s\F}\Ber$ and $\ ^{int}\Omega^{p-n}_{s\F}$  with sections of some vector bundles $A_p$ and $B_p$ over $M_{red}$. It is fairly straightforward to establish an isomorphism of $A_p$ and $B_p$ with the help of the splitting. In particular there is a $C^{\infty}$ isomorphism $\Ber=\ ^{int}\Omega^{-n}_{s\F}$. 

One can think about $\Omega^{0}_{s\F}$ as of a space of functions on a supermanifold $sM$. Then a space of sections $Ber(sM)$ coincide with $\Omega^{n}_{s\F}\Ber)$. It elements   can be integrated over $sM$. The integral  defines a pairing $(.,.)_s:\Omega^{\bullet}_{s\F}\otimes \Omega^{\bullet}_{s\F}\Ber)\overset{\int<.,.>_{s}}\rightarrow \mathbb{C}$, which is nondegenerate.

An element $a\in \ ^{int}\Omega^{0}_{s\F}$ defines a functional $f\rightarrow \int_{M} a f$ ($f\in C^{\infty}(sM)$). We may think about it as of an integral $\int_{sM} fi(a)$ , where $i(a)\in \Omega^{n}_{s\F}\Ber$. Such interpretation of the integral uniquely specifies map $i$. Since $\Omega^{n}_{s\F}$ is invertible the induced isomorphism $i:\Ber=\ ^{int}\Omega^{-n}_{s\F}$  is compatible with $\dbar$ (use pairings $(.,.),(.,.)_s$ to check this).

\end{proof}
Suppose that a super CR-manifold $M$ is CR embedded into a complex super manifold $N$. Denote by $J$ an operator of complex structure in tangent bundle $TN$. We assume 

$TM+JTM=TN_{|M}$. Denote  by $Ber_{\mathbb{C}}(N)$ the complex Berezinian of  $N$. An easy local computation shows that $Ber_{\mathbb{C}}(N)_{|M}$ is isomorphic to $\Ber$. Suppose  $N$ is a Calabi-Yau manifold , i.e. it admits a global nonvanishing section $vol$ of $Ber_{\mathbb{C}}(N)$. A  restriction  of $vol$ on $M$ defined a global CR-holomorphic section of $\Ber$. An isomorphism of Proposition \ref{P:fdfdhw}  provides a $\dbar$ closed section of $\ ^{int}\Omega^{-n}_{s\F}\subset \ ^{int}\Omega^{-n}_{\F}$. 

\begin{remark} The proof Proposition \ref{P:fdfdhw} parallels with the proof of Serre duality in super case given in ref. [HW]. Haske and  Wells used sheaf-theoretic description of a complex supermanifold, which significant simplify the argument . The main simplification comes from the local Poincare lemma, which is absent is CR-case.
\end{remark}

It is  worthwhile to mention that there is no canonical ($Aut_{\F}$ or $Out_{\F}$ equivariant)   map $ \Omega^{\bullet}_{\F}\Ber\rightarrow \ ^{int}\Omega^{\bullet}_{\F}[-n]$. This seems to be one of fundamental distinctions of purely even and super cases.

We think the reason is that the only known construction of this map is through the intermediate complex $\Omega^{\bullet}_{s\F}\Ber$. As have have already mentioned in remark (\ref{E:fdfdsdfv}) the complex $\Omega^{\bullet}_{s\F}$ carries only A$\ity$ action of $Aut_{\F}$.

A real line bundle $Ber$ on a complex super manifold  is a tensor product $Ber_{\mathbb C}\otimes \overline{Ber}_{\mathbb C}$, where $Ber_{\mathbb C}$ is a holomorphic Berezinian.

A decomposition $T^{\mathbb{C}}={\cal T}=\overline{{\cal T}}$ implies that

\begin{equation}
\begin{split}
&^{int}\Omega^{-k}=Ber\otimes \\
&\Lambda^{k}({\cal T})=\bigoplus_{i+j=k} Ber_{\mathbb C}\otimes \Lambda^{i}({\cal T}_{\mathbb C})\otimes \overline{Ber}_{\mathbb C} \otimes \Lambda^{j}(\overline{{\cal T}}_{\mathbb C})=\bigoplus_{i+j=k}\ ^{int}\Omega^{-i,-j}
\end{split}
\end{equation}

\begin{remark}\label{E:hhghssa}
Suppose $M$ is a complex  $n$-dimensional manifold, $E$ is an $k$-dimensional vector bundle. 
On the total space $ \Pi E$ there is a canonical section of $c_{\Pi E} \in Ber_{\mathbb{C}}\otimes \Lambda^{n}({\cal T})$. 
In local odd fiberwise  coordinates $\theta_i$ it is equal to 
\begin{equation}\label{E:jsdfdmdas}
c_{\Pi V}=\theta _1\dots \theta _k d\theta _1\wedge \dots \wedge d \theta _k.
\end{equation}

\end{remark}
\begin{proposition}\label{P:kahedy}
The forms $c_{\Pi V}$ and $\bar c_{\Pi V}$ are $\dbar$-closed.
\end{proposition}
\begin{proof}
Since the formula (\ref{E:jsdfdmdas}) does not depend on a choice of coordinates on $M$ one can do a local computation, which is trivial.
\end{proof}

\section{Acknowledgments}
The author would like to thank P. Candelas, P.Deligne, M.Kontsevich, L. Maison, A.S. Schwarz for useful comments and discussions. The note was written while the author was staying at Mittag-Leffler Institut, Max Plank Institute for Mathematics and Institute for Advanced Study. The author is grateful to these institutions for kind  hospitality and support. After the preprint version of this paper [Mov] had been published, the author learned about the work by L.J.Mason, D.Skinner [MS] where they treat a similar problem.

\end{document}